\documentclass[useAMS,usenatbib]{mn2e}

\usepackage{amssymb, amsmath}
\usepackage{graphicx}
\usepackage{mathtools}
\usepackage{longtable}
\newcommand{\apj}{ApJ}

\newcommand{\apjl}{ApJL}
\newcommand{\aap}{A{\&}A}

\newcommand{\mnras}{MNRAS}

\newcommand{\pasj}{PASJ}

\title[Warped Gas Disks]{On the Formation of Warped Gas Disks in Galaxies}

\author[Haan \& Braun]{S. Haan$^{1}$ \& R. Braun$^{2}$\\
$^{1}$CSIRO Astronomy and Space Science, ATNF, PO Box 76, Epping 1710, Australia\\
$^{2}$SKA Organisation, Jodrell Bank Observatory, Lower Withington, Macclesfield, Cheshire, SK11 9DL, UK}

\begin{document}


\pagerange{\pageref{firstpage}--\pageref{lastpage}} \pubyear{2014}

\maketitle

\label{firstpage}

\begin{abstract}
We consider the most commonly occurring circumstances which apply to galaxies, namely membership in galaxy groups of about $10^{13}h^{-1}~M_\odot$ total mass, and estimate the accompanying physical conditions of intergalactic medium (IGM) density and the relative galaxy-IGM space velocity. We then investigate the dynamical consequences of such a typical galaxy-IGM interaction on a rotating gaseous disk within the galaxy potential. We find that the rotating outer disk is systematically distorted into a characteristic ``warp'' morphology, of the type that has been well-documented in the majority of well-studied nearby systems. The distortion is established rapidly, within two rotation periods, and is long-lived, surviving for at least ten. A second consequence of the interaction is the formation of a one arm retrograde spiral wave pattern that propagates in the disk. We suggest that the ubiquity of the warp phenomenon might be used to reconstruct both the IGM density profile and individual member orbits within galaxy groups.
\end{abstract}


\section{Introduction}

Although historically is was customary to consider galaxy evolution to occur in relative isolation, there has been a growing realisation that interactions amongst galaxies and with the broader inter-galactic environment play an important role in determining both the evolution and the ultimate fate of individual galaxies. Two assumptions in particular still color the intuition of many researchers, namely (1) that most normal galaxies can be considered to be essentially at rest with respect to their environment, and (2) that the typical intergalactic densities are so low that they can be neglected. But are those assumptions justified?\par
Comparative studies of large observational surveys of galaxy populations in the local universe with cosmological numerical simulations have established that the majority of galaxies occur in groups with a median total mass of about $10^{13}h^{-1}$ M$_\odot$, and neither in isolation nor in massive clusters \citep{eke05,nur13}. Such mass concentrations will be self-gravitating out to radii that enclose a mean mass density of about, $<\rho> = 200 \rho_{crit}$, for a critical density, $\rho_{crit} = 3 H^2_0/(8 \pi G)$, with the Hubble constant, $H_0$, so that $M_{200} = 200 \rho_{crit} (4\pi/3) r^3_{200}$. The corresponding median group radius is therefore about $r_{200} = 350 h^{-1}$ kpc.\par
It has also been demonstrated that the characteristic median projected galaxy separation within such groups is, $<R> = 0.13 h^{-1}$ Mpc, and the line-of-sight velocity dispersion is $\sigma_V = 150$ km~s$^{-1}$ \citep{nur13}. The corresponding physical radius and space velocity for an isotropic distribution is $<r> = (3/2)^{0.5}<R> = 0.16 h^{-1}$ Mpc and $\sigma_{3d} = 3^{0.5}\sigma_V = 260$ km~s$^{-1}$. The intergalactic medium densities within galaxy group environments may also conform to a ``universal'' radial profile. This was first suggested by \citet{nav96} for the dark matter profile and has been verified in many subsequent cosmological simulations. More recently, \citet{dav10} have suggested that a characteristic baryonic mass over-density $\rho_c/\bar{\rho} = 120 \pm 0.2$ dex occurs at the radius, $r_{200}$, of self-gravitating halos of any mass, with $\bar{\rho}$ as the mean baryon density in the Universe. At intermediate radii, both an isothermal and an NFW density profile \citep{nav96} predict a baryonic density that varies as $(r/r_{200})^{-2}$. This allows estimation of the corresponding volume density from,
\begin{equation}
 n_H(r) = {\rho(r) \over \mu m_H} = {\rho_c \over \bar{\rho}} \bigg({r \over r_{200}}\bigg)^{-2} {\Omega_b 3 H^2_0 \over \mu m_H 8 \pi G}.
\end{equation}
With $\Omega_b = 0.049$, $H_0 = 67.3$ km/s/Mpc \citep{ade13} and a mean atomic mass, $\mu = 1.4$, this yields a characteristic IGM density, $n_H(<r>) = 1 \times 10^{-4}$ cm$^{-3}$.\par
The orbital timescale of galaxies in a typical group environment can be estimated from their radii and velocity dispersion as, $P \approx 2 \pi r_{200} / \sigma_{3d} = 3.8$ Gyr. Similar ambient conditions of IGM density and relative galaxy-IGM space velocities might then be expected to apply for some fraction of that orbital timescale, say 1 - 2 Gyr. So, rather than being at rest in ``empty'' space, the vast majority of galaxies have relative velocities of $\sigma_{3d} = 260$ km~s$^{-1}$ through an IGM of density $n_H = 1 \times 10^{-4}$ cm$^{-3}$ and such interactions can build up ``coherently'' for time intervals of several Gyr. What might the consequences of such interactions be?\par 
In an earlier paper \citep{haa13} we considered whether a detectable kinematic signature of an ``instantaneous'' ram pressure interaction might be imparted to the 21cm HI velocity fields of nearby galaxies. Our analysis demonstrated that a distinctive kinematic perturbation is induced that allows the ram pressure components both perpendicular and parallel to a galaxy disk to be deduced from sensitive HI emission observations at large radii. The fact that these kinematic perturbations of the velocity field are concentrated in distinct Fourier modes, m=0 for the perpendicular interaction, and m=2 for that parallel to the disk meant that they could be robustly distinguished from the vast majority of other kinematic perturbations that tend to concentrate in the m=1 Fourier mode. One of the most important m=1 kinematic perturbations is the ``warp'' phenomenon.\par 
While it has been empirically determined that the majority of disk galaxies display a significant degree of warping of the gaseous disk at large radii and that there are well-documented systematic patterns of warp morphology and kinematics \citep[see e.g.][]{Bri90, Kam92, Joz07, Kru07, Kam13}, there is still no satisfactory explanation for the origin and persistence of this phenomenon. 
In particular the possibility that warps are caused by accretion remains as the most plausible explanation, which can either result in a  reorientation of the halo by the accretion of matter with mis-aligned angular momentum \citep{Ost89, Jia99} or an accretion flow which intersects a galactic disk \citep{Kah59, May81, Rev01, Lop02, San06}. Indeed \citet{Lop08} have sought systematically aligned warps within galaxy samples as a probe of large scale accretion signatures. Several other hypothesis for warp formation have been suggested \citep[see for more details][]{San06}, such as magnetic fields \citep[e.g.][]{Bat98}, interaction with satellites \citep[e.g.][]{Hua97} with possible amplification due to the halo \citep{Wei98}, vertical resonance \citep[][]{Gri02} or bending instabilities \citep{Rev04}.\par
In this study we consider what the long term consequence might be of galaxy propagation though an IGM of significant density. By tracking the cumulative effect of the ram pressure interaction with a rotating gaseous disk in a realistic potential we demonstrate that a natural consequence of this interaction is significant warping of the outer disk. The warp phenomenon sets in within about two rotational periods and persists for more than ten. We begin with a brief consideration of the ram pressure forces and test particle responses in \S 2 and then consider more extensive numerical simulations in \S 3. The results are presented and briefly discussed in \S 4.

\section{Model of ram pressure impact on gas orbits}
\label{sec:orbit}
The gas component of a galaxy's disk responds in a measurable way to ram forces, which can lead to a significant change of the gas velocity field on less than one orbital timescale \citep[see][]{haa13}, and over longer time-scales, will eventually significantly alter the orbits of gas clouds. Here we test how ram forces can effectively change the orbital configuration of gas clouds over several dynamical timescales ($>$1Gyr). The force acting on a gas cloud is $\mathbf{F=F_{grav} + F_{Ram}}$, where $\mathbf{F_{grav}}=-\nabla \Phi_{grav}$ is the force in the gravitational potential of a galaxy  and $\mathbf{F_{Ram}}$ the ram force due to the relative velocity between disk galaxy and its intergalactic environment.
The ram pressure is defined as $P_{ram}\propto \rho_{ISM}\;(\mathbf{v_{gal}-v_{IGM}})^2$  where $\mathbf{v_{gal}}$ is the velocity vector of the galaxy and $\mathbf{v_{ICM}}$ the velocity vector of the IGM \citep{Gun72}.\par	
In a first test we calculate the trajectory of a test particle in a halo potential which is subject to a drag force due to ram pressure. The gravitational potential is calculated for a spherical mass distribution of an isothermal sphere, which is equivalent to a logarithmic potential and a gravitational force $\mathbf{F}_{grav}=v_{max}^2 \mathbf{R}/R^2$
resulting in a flat rotation curve with $v_{max}=150$~km~s$^{-1}$, which is a fair approximation of gas rotation in the outer disk of most galaxies. We will discuss more realistic gravitational potentials in \S~\ref{sec:sim}, which takes into account the response of gas clouds on all spatial scales due to the combination of dark matter and stellar gravitational potential. The acceleration of a gas cloud due to ram pressure is given as 
\begin{equation}
\mathbf{a_{ram}}=\frac{c(R)\;(\mathbf{v_{gal}}-\mathbf{v_{IGM}})\;|\mathbf{v_{gal}}-\mathbf{v_{IGM}}|}{1+\frac{\rho_{ISM}}{\rho_{IGM}}}
\end{equation}
The coefficient $c$ takes into account the sum of all factors associated with the collision process between ISM and IGM, i.e. the exposed surface area of gas clouds towards the ram wind, the gas cloud mass, and possible dissipational terms depending on nature of the  collision between ISM and IGM \citep[see for more details][]{haa13}.\par
\begin{figure}
\begin{center}
\includegraphics[scale=0.45]{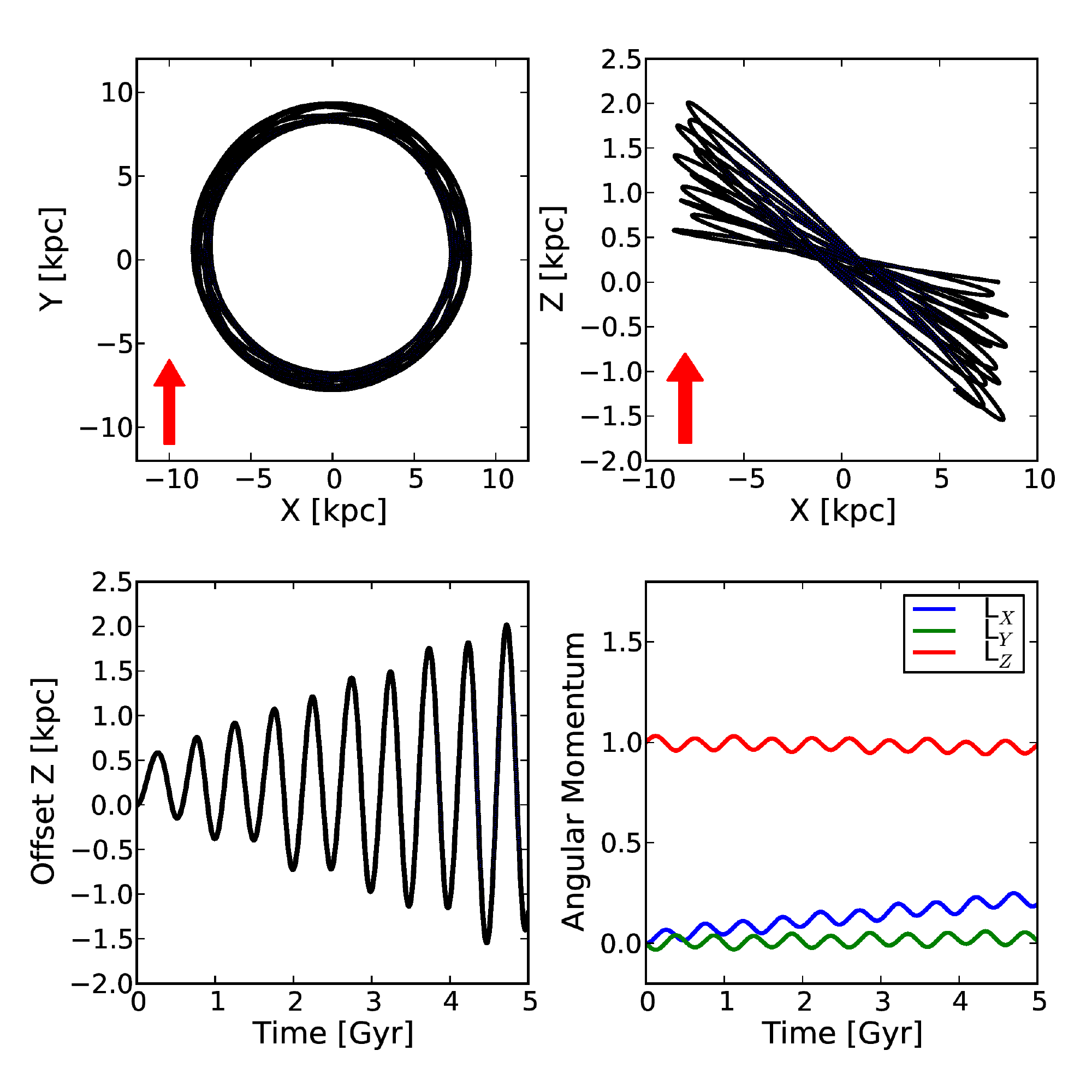}
\caption{Top panels: Trajectory of a counter-rotating test particle as function of position and time in a spherical logarithmic potential with a ram force in the (Y,Z) directions with an inclination angle of 45~degree (red arrow). The bottom left panel displays the offset in vertical direction as a function of time, while the bottom right panel shows the normalized decomposition of the angular momentum vector $\mathbf{L}=\mathbf{R \times v}$. } 
\label{fig:orbit}
\end{center}
\end{figure}

In Fig.~\ref{fig:orbit} we show the trajectory of a gas cloud as function of position and time. At $t=0$, the test particle is placed at a radius of 8~kpc with circular motion in the $X-Y$ plane. Here we apply a ram wind component with an inclination of 45~degree between horizontal plane and vertical axis. We find that the ram force causes a variation in the elongation of the orbit in the horizontal plane (change in $r=\sqrt{X^2+Y^2}$) while increasing the inclination of the orbit in the vertical direction as function of orbital timescale. The external torque due to the ram force changes the direction of the angular momentum vector of the gas cloud, $\mathbf{L}/m=\mathbf{R \times v}$. These results show that even a moderate ram pressure force ($F_{ram}<F_{grav}$) leads naturally to a significant change of the inclination of a gas cloud's orbit over a few dynamical time-scales.

\section{Numerical Simulations}
\label{sec:sim}
To test how a gas disk of a galaxy responds to a ram pressure field, we simulate the orbital paths of gas clouds in a static galactic potential.  While the interaction of gas clouds via merging and dissipation is certainly very important for many galaxy evolutionary processes, in particular on smaller scales, we assume that the main factor for the global  dynamics of a gas disk is the underlying gravitational potential and interaction processes with its environment. Therefore our model does not rely on any hydrodynamical assumptions as usually applied in smooth particle hydrodynamics (SPH) or sticky particle simulations.\par
The gravitational potential in our model is given by the stellar disk and a spherical halo component.
The stellar disk potential can be approximated by the Miyamoto \& Nagai potential \citep[see][]{Miy75, Bin87}: 
\begin{equation}
\Phi_{disk}(r,z)=\frac{G\;M_{disk}}{\sqrt{r^2 + (a_{disk} + \sqrt{z^2 + b^2_{disk}})^2}}
\end{equation}
where $M_{disk}=1.5 \times 10^{10}$~M$_\odot$, $a_{disk}=1$~kpc, and $b_{disk}=100$~pc.
The gravitational potential of the spherical halo is given as \citep{Bin87}
\begin{equation}
\Phi_{halo}(R)=\frac{1}{2}v_0^2\mathrm{ln}(R_c^2 + R^2 + z^2)
\end{equation}
with $v_0=150$~km~s$^{-1}$ and $R_c=2$~kpc.\par
\begin{figure}
\begin{center}
\includegraphics[scale=0.5]{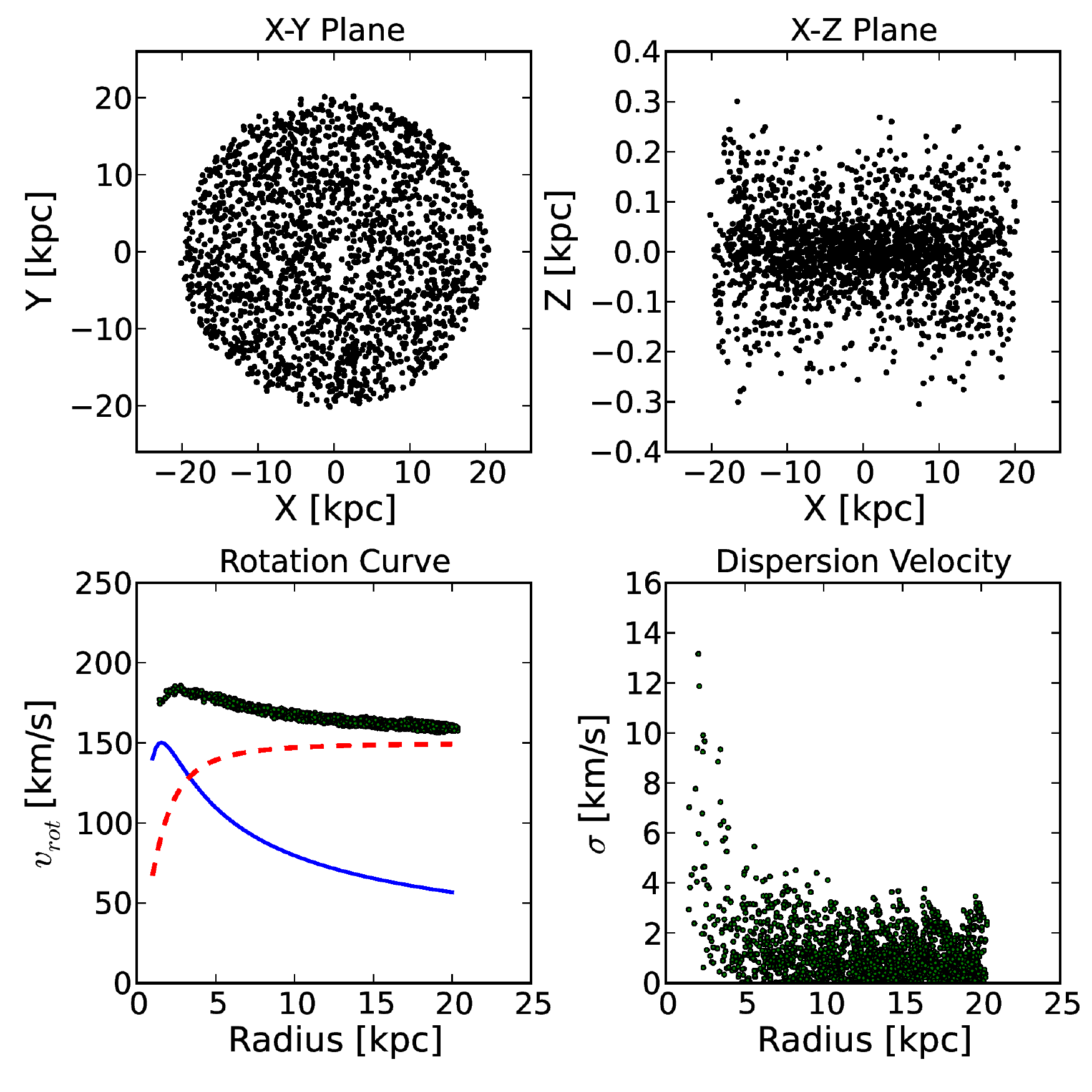}
\caption{Distribution of test gas clouds in the $(X,Y)$-plane (top left) and $(X,Z)$-plane (top right) after two dynamical time-scales ($\sim$1.0~Gyrs) in the combined gravitational potential of a disk and spherical halo. The bottom left panel shows the circular rotation velocities (black points) as function of radius, which can be decomposed into a stellar disk (blue line) and a spherical halo component (dashed red line). The dispersion velocity measured from the velocity component in the Z-direction is shown in the bottom right panel.} 
\label{fig:init}
\end{center}
\end{figure}

The self-gravity of the gas component is neglected since the gas typically contributes less than 5\% to the total gravitational mass of a galaxy. The collisionless stellar disc is not directly affected by ram pressure and hence shows no disturbed kinematics or distribution \citep[see][]{Kro08}. The gas clouds are uniformly distributed over a disk with a radius of 20~kpc (typical HI disk) to trace the response of the gas due to the combined forces of the gravitation potential and ram pressure. The initial velocities in the (X,Y) plane are defined as circular orbits corresponding to the gravitational potential of the galaxy.  For each particle a random dispersion term is added to the (X,Y,Z) velocities using a Gaussian distribution with $\sigma$ = 4~km~s$^{-1}$ and we let the gas further relax over two dynamical time-scales ($\sim$1~Gyr).  Fig.~\ref{fig:init} shows the gas distribution and rotation curve after this initialization run. Then we add the ram wind component and let the gas evolve over 4~Gyrs ($\sim$8 dynamical time-scales). 
We have chosen an average density ratio of ISM/IGM of 500:1 and a ram wind of 300~km~s$^{-1}$, which is a typical velocity of field and group galaxies relative to the rest-frame of the IGM. These ram pressure parameters ensure that the ratio of $F_{grav}/F_{ram}<1$ on all scale lengths and hence does not lead to a stripping of gas.
We have simulated the impact of the ram wind for four wind directions, a) perpendicular to the disk, b) parallel to the disk, c) 45~degree inclined to the disk in (X,Z)-direction, and d) 45~degree inclined to the disk in (Y,Z)-direction. In a first run we assume that all gas clouds have the same density and hence experience the same effective ram force. In subsequent runs we test more realistic gas density profiles with a transition from clumpy to diffuse gas with increasing radius.

\section{Results and Conclusions}
\begin{figure}
\begin{center}
\includegraphics[scale=0.44]{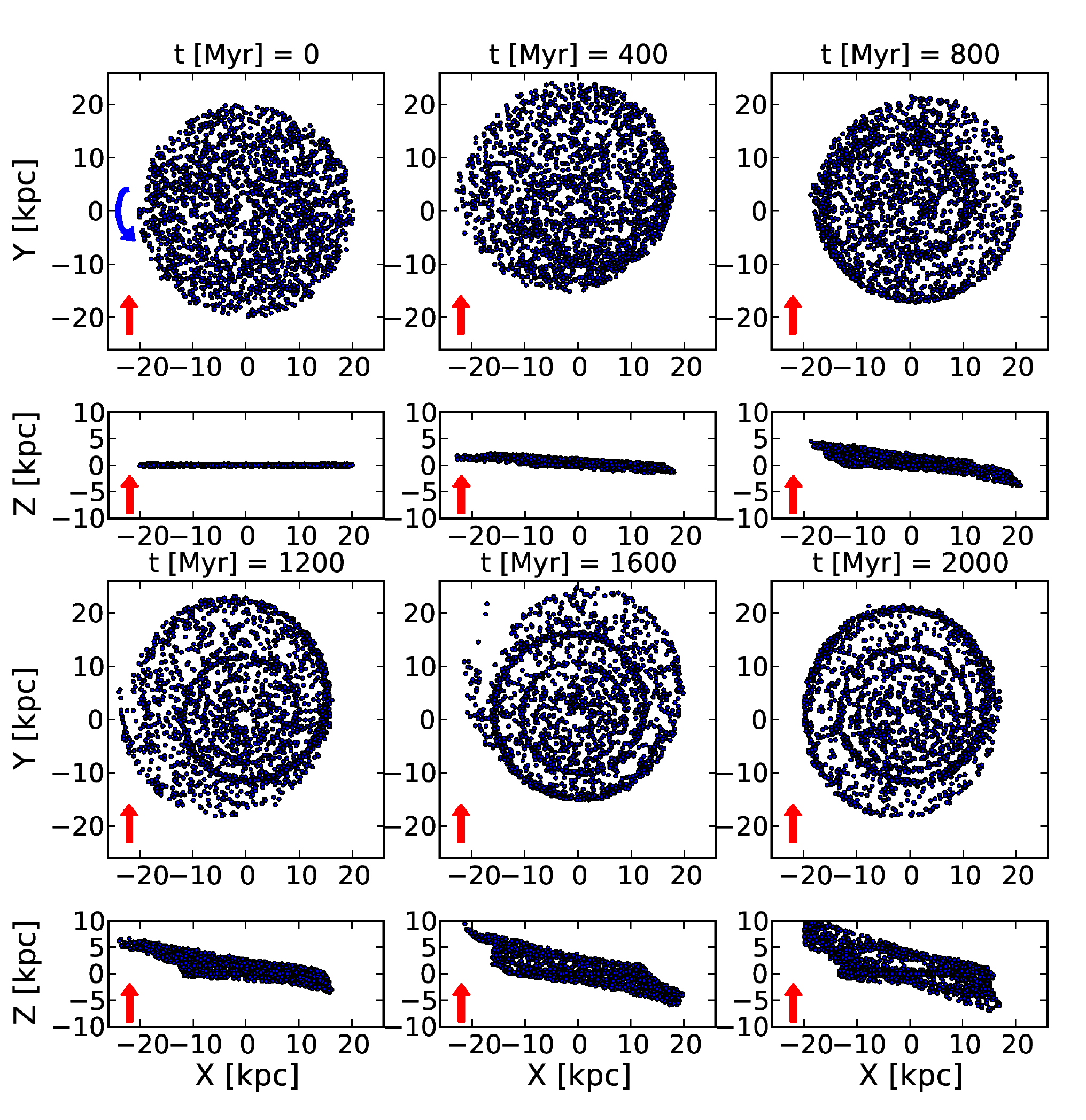}
\caption{The gas distribution as function of time under ram pressure (45~degree angle between X and Z axis, the red arrow indicates the wind direction) in the combined gravitational potential of disk and spherical halo. All gas clouds have the same density with a uniform effective ram pressure and the galaxy is rotating in counter-clockwise direction (blue arrow).} 
\label{fig:time}
\end{center}
\end{figure}

\begin{figure}
\begin{center}
\includegraphics[scale=0.54]{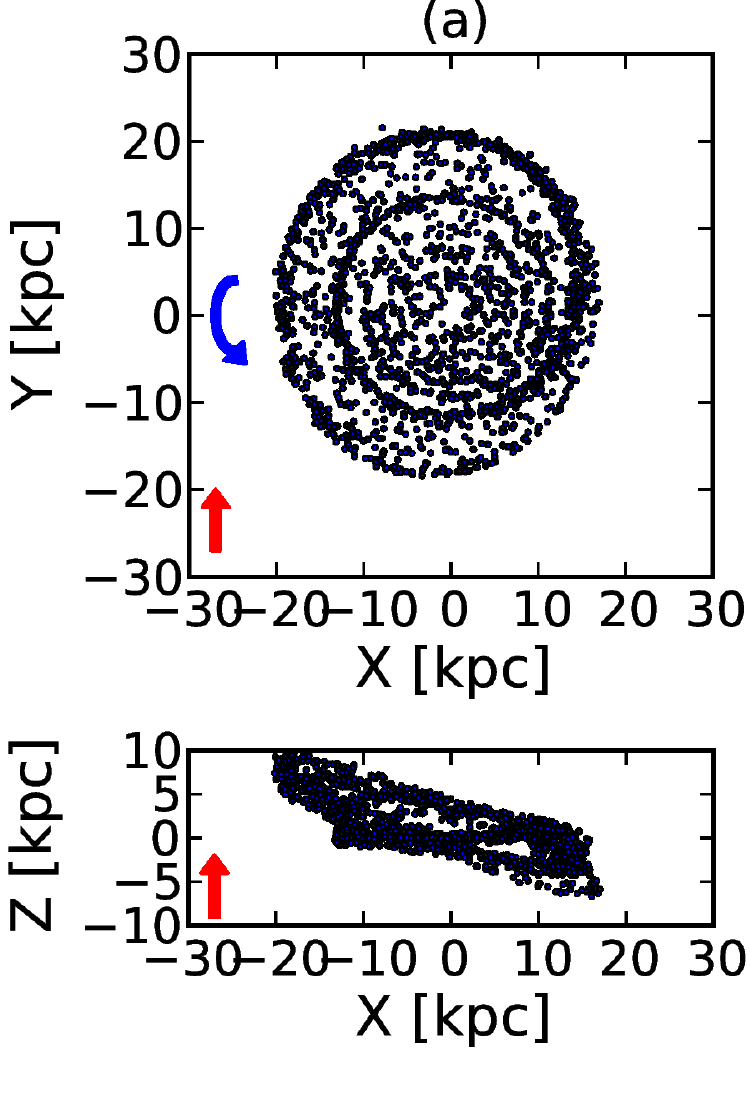}
\includegraphics[scale=0.54]{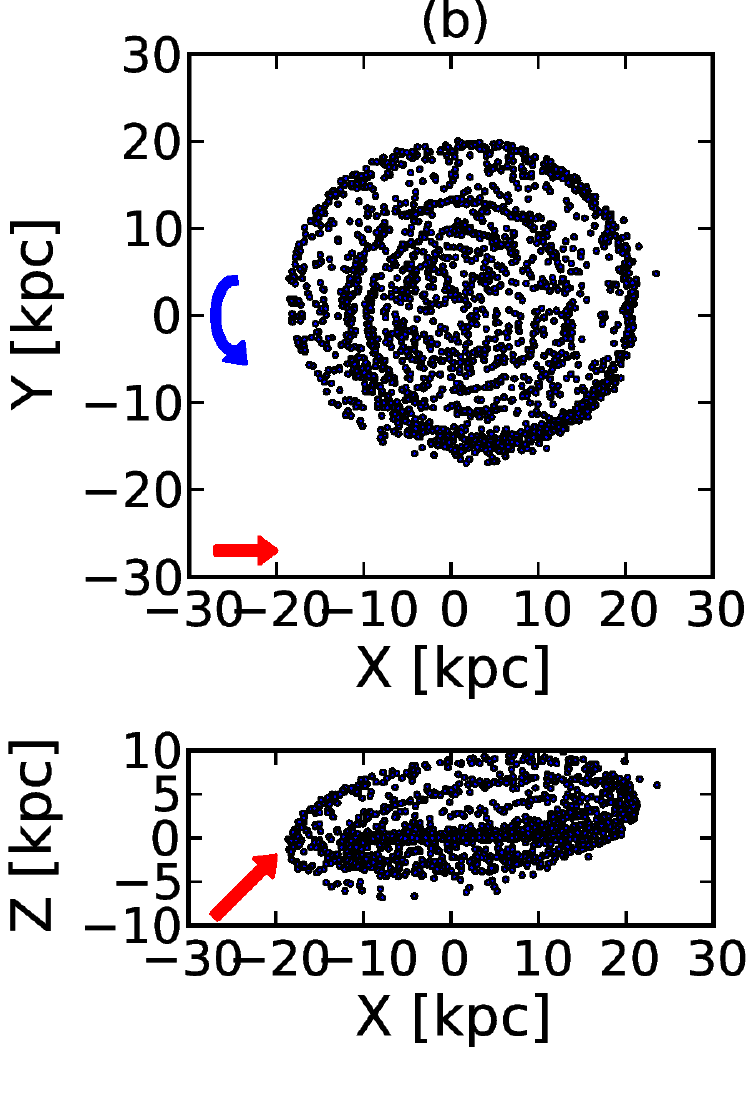}
\includegraphics[scale=0.54]{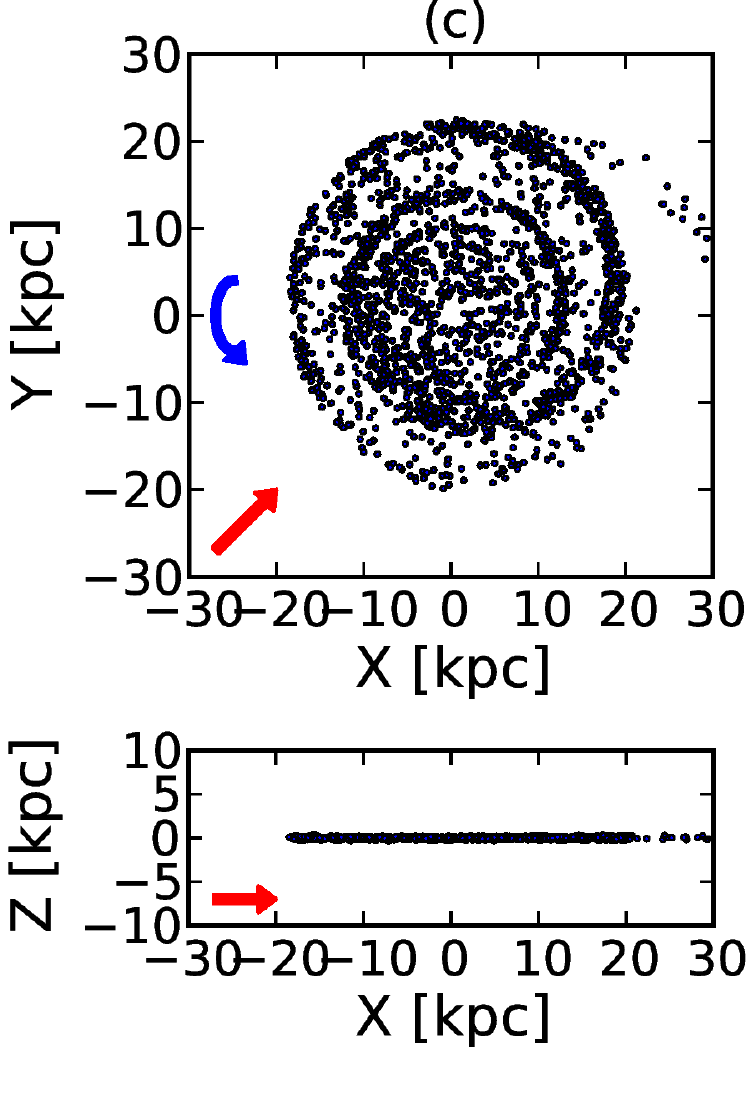}
\includegraphics[scale=0.54]{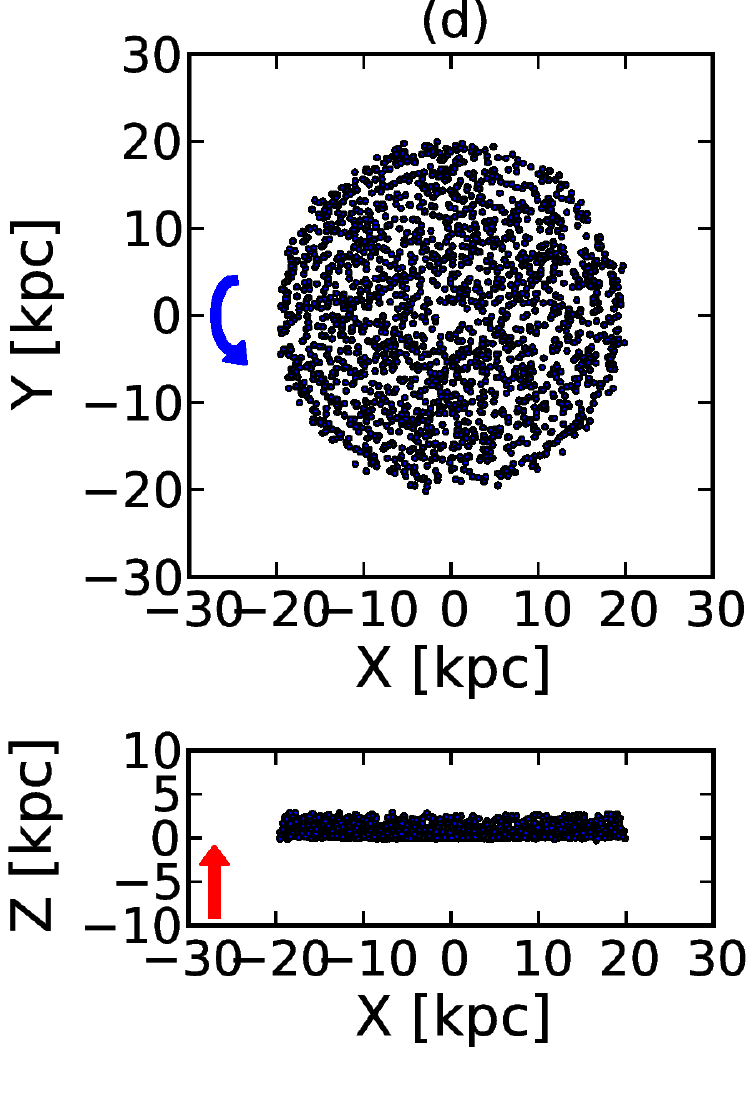}
\caption{Comparison of the gas distribution after 4~Gyrs under ram pressure with a ram wind of 300~km~s$^{-1}$ for four different directions (red arrow): (a) in (Y,Z) direction with 45~degree inclination to disk (top left), (b) in (X,Z) direction with 45~degree inclination to disk (top right), (c) in (X,Y) direction (parallel to disk, bottom left), and (d) in Z-direction (perpendicular to disk, bottom right).} 
\label{fig:direction}
\end{center}
\end{figure}

Fig.~\ref{fig:time} displays the gas distribution as function of time for a gas disk subjected to a uniform ram pressure. We find that the ram force leads to a significant change of the orbital paths of the gas clouds, but with very different orbital alignments as function of ram wind angle (see Fig.~\ref{fig:direction}): A ram wind parallel to the disk causes only a periodic variation in the plane direction as function of time (periodic lopsidedness), while a perpendicular ram wind leads only to a small displacement in the vertical direction of the disk along the ram wind angle, inducing an additional ``flaring'' or a minor ``U-shaped'' warp \citep[non-axisymmetric flares, see also][]{Lop09}. Only a combination of perpendicular and parallel ram forces, namely a ram wind that is moderately inclined to the disk, leads to a significantly ``S-shaped'' warped disk as shown in Fig.~\ref{fig:direction} (see also Fig.~S1 and movie in supplementary material). This result can be explained in a simple picture: The orbits of gas clouds on one side of the galaxy are elongated due to the ram component in the (X,Y)-direction while on the other side they are shortened. This leads to an asymmetry where gas clouds are exposed for a longer time and to a more effective ram wind in the z-direction due to the smaller gravitational forces at larger radii on one half of the orbit than on the other. Integrated over the entire orbital path, this causes a change in the inclination angle and a warped disk evolves.  Our results show that the warp direction depends on the direction of the galaxy's motion through the IGM and whether the galaxy is rotating in the clockwise or counter-clockwise direction, which results in a distortion which has the opposite sign in the Z-direction (see also Fig.~S2 in the supplementary material). The formation of warped structure is in agreement with studies of torques exerted by accretion flows \citep[e.g.][]{Lop02, San06}, and our results corroborate the picture that warps in galactic disks are formed by the interaction with their surrounding material.\par 
The radius at which the disk becomes warped depends not only on the gravitational potential but also significantly on the state of the neutral ISM, which goes through a phase transition from dense clumpy clouds (with possible molecular gas cores) towards lower density clouds and diffuse gas \citep[see][]{haa13}. 
We have compared the geometry of the final warped gas disk after 4~Gyrs for three different density profiles: (a) uniform, (b) exponential, and (c) hyperbolic with a transition radius at 15~kpc (see Fig.~S3 in the supplementary material). 
In all cases the final gas distribution and velocities show that the ram wind induces an ``S-shaped'' warped  gas disk for a moderately inclined ram wind angle. However, the radial scale at which the inclination of the disk changes depends on the density properties of the ISM.\par
Besides the warped morphology, we find that a ram wind component parallel to the plane of the galaxy induces a one-arm spiral pattern that slowly evolves with time in a retrograde direction. The nature of this structure can be interpreted as a wave phenomenon due to the combination of the underlying gravitational potential and the ram wind direction in the plane of the galaxy. \cite{Tos94} has shown a similar structure evolving under ram pressure, but was restricted only to the plane of the galaxy and no vertical component was included. If the ram wind also has a vertical component, as is the case in our study, the retrograde spiral in the disk evolves into a warped retrograde helix structure along the vertical axis as shown in Fig.~\ref{fig:direction} and in the simulation movie (see supplementary material). Moreover, we find that the ram pressure induced ``instantaneous'' non-circular motions ($\sim 10 - 50$~km~s$^{-1}$ over less than one orbit) are in the same range as measured in our kinematic ram pressure study \citep{haa13}, which suggests that both, the warped geometry and the non-circular motions that are not due to warps, require similar ram pressure properties (IGM density and relative velocity of galaxy to IGM).\par
This study demonstrates that the measured typical motion of a disk galaxy relative to its median intergalactic environment can lead to a significant change of the morphology of the gas distribution, characterized by a ``S-shaped'' warped, lopsided disk and a one arm retrograde helix structure. Given the results of our previous kinematic ram pressure study \citep{haa13} and the fact that most galaxies show these structures, we suggest that the interaction between galaxies and their surrounding IGM are the main drivers for both non-planar and non-circular motions in the outer gas disks of galaxies. Having recognised the cause of these distortions, one can use the ubiquity of these phenomena to infer the physical attributes of galaxy groups. With multiple probes of the same environment, each galaxy within a group can be used to infer the likely IGM density profile, the galaxy space velocity and a likely orbital history.

\section*{Acknowledgements}
The authors wish to thank the anonymous referee for a careful review of the manuscript and many valuable suggestions to improve this paper. The authors are grateful to Romeel Dav\'e for helpful communication on dark matter halos in numerical simulations.



\clearpage

%
%
%
%
\label{lastpage} 


\begin{figure*}[!h]
\begin{center}
\includegraphics[scale=0.5]{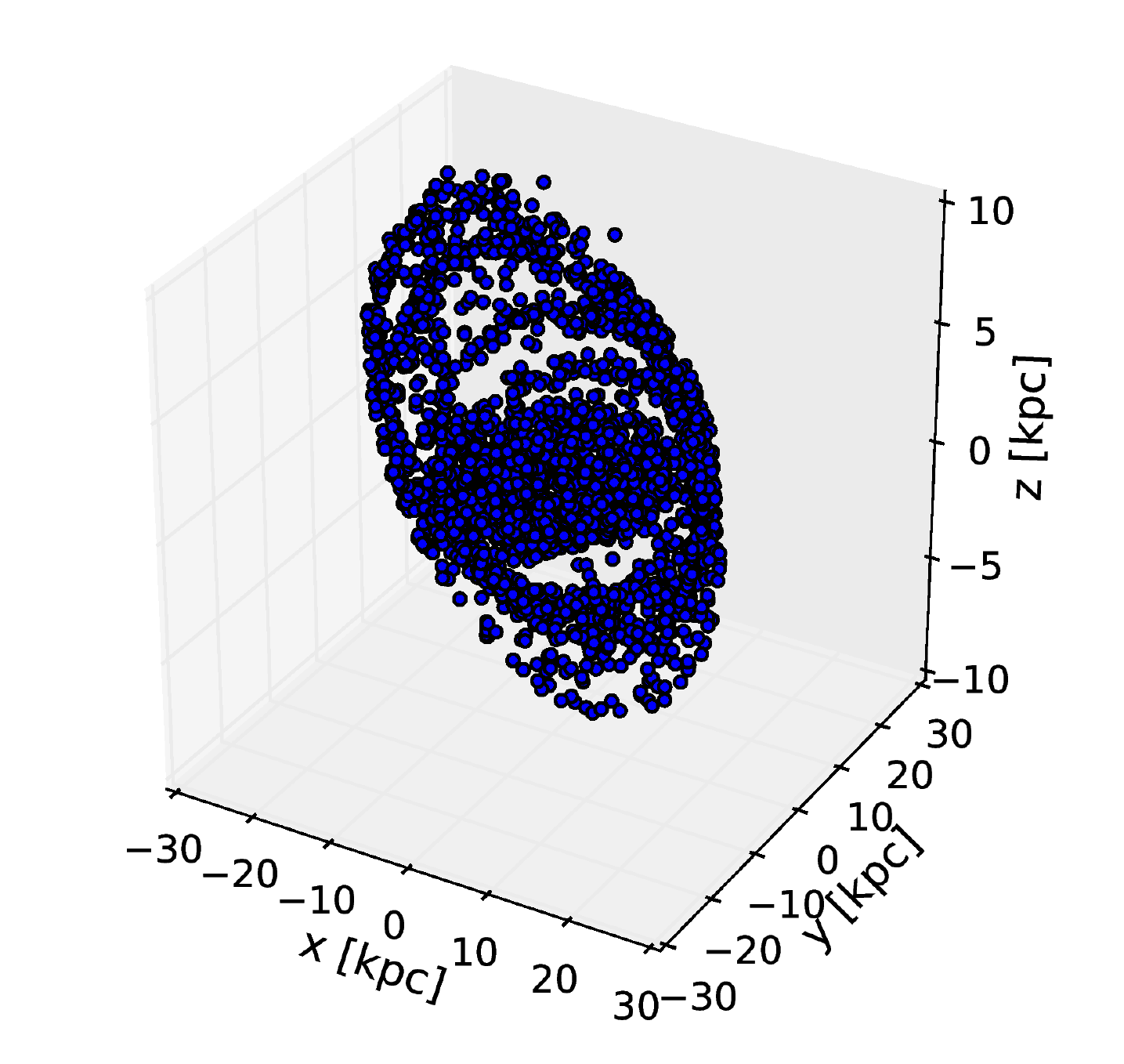}
\includegraphics[scale=0.4]{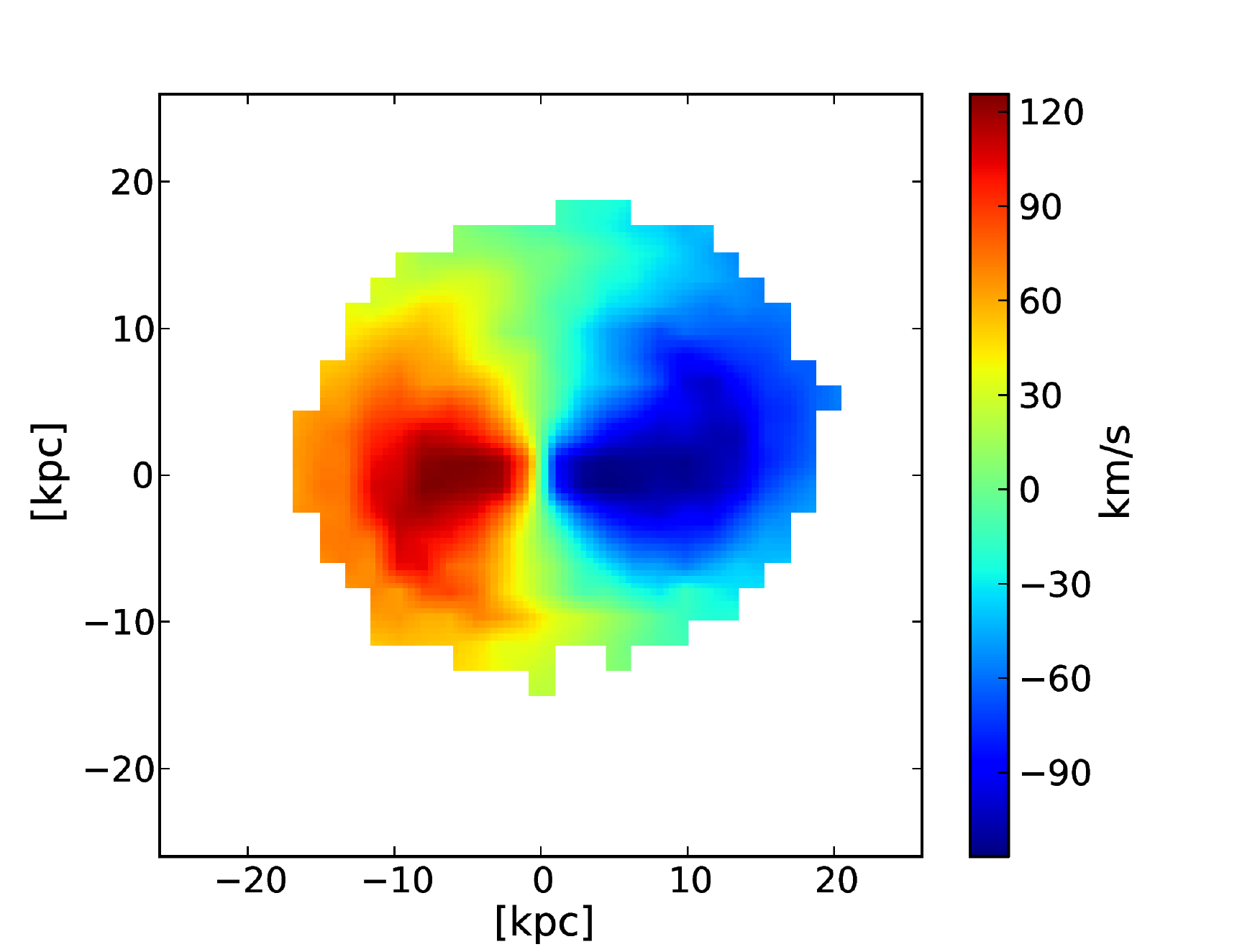}\\
\textbf{Figure~S1}: The gas distribution (left panel) and the line-of-sight velocity fields (right panel) as observed on the sky with an inclination of 45~degree of the final warped disk after 4~Gyrs (ram properties and disk geometry are the same as in Fig.~3).
\end{center}
\end{figure*}

\begin{figure*}[!h]
\begin{center}
\includegraphics[scale=0.8]{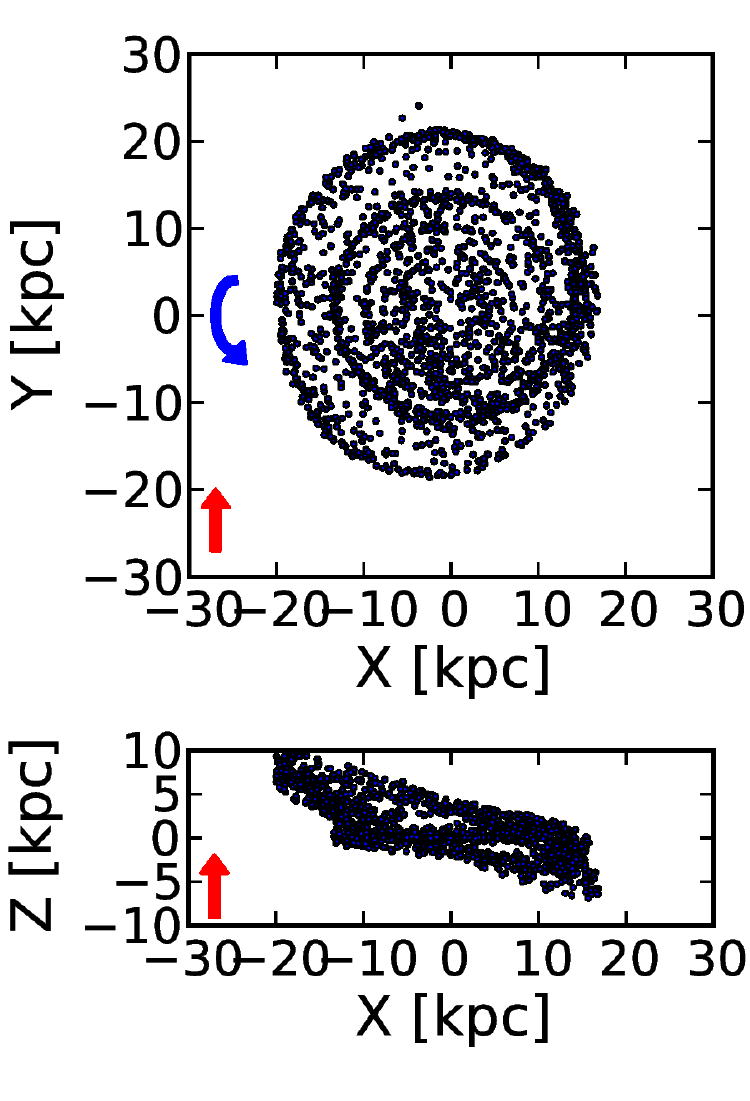}
\includegraphics[scale=0.8]{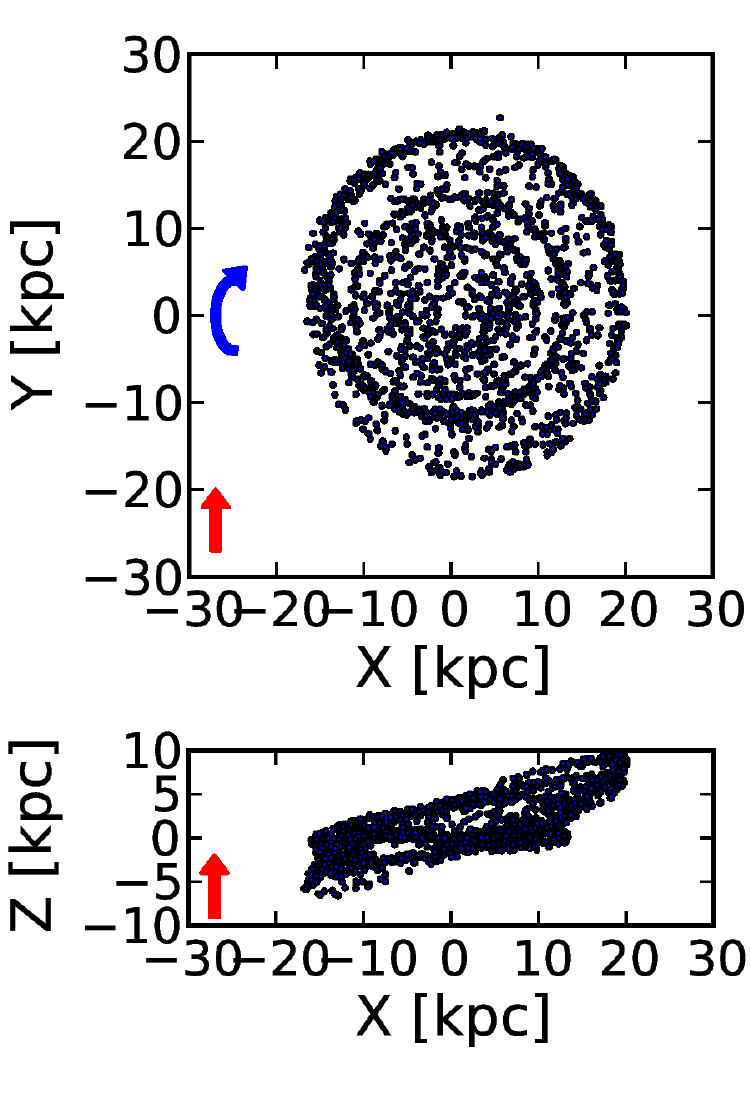}\\
\textbf{Figure~S2}: Comparison of the geometry of the final warped gas disk after 4~Gyrs under a moderate inclined ram wind between counter-clockwise (left panel) and clockwise rotation (right panel). Ram properties and disk geometry are the same as in Fig.~3. 
\end{center}
\end{figure*}

\begin{figure*}[!h]
\begin{center}
\includegraphics[scale=0.54]{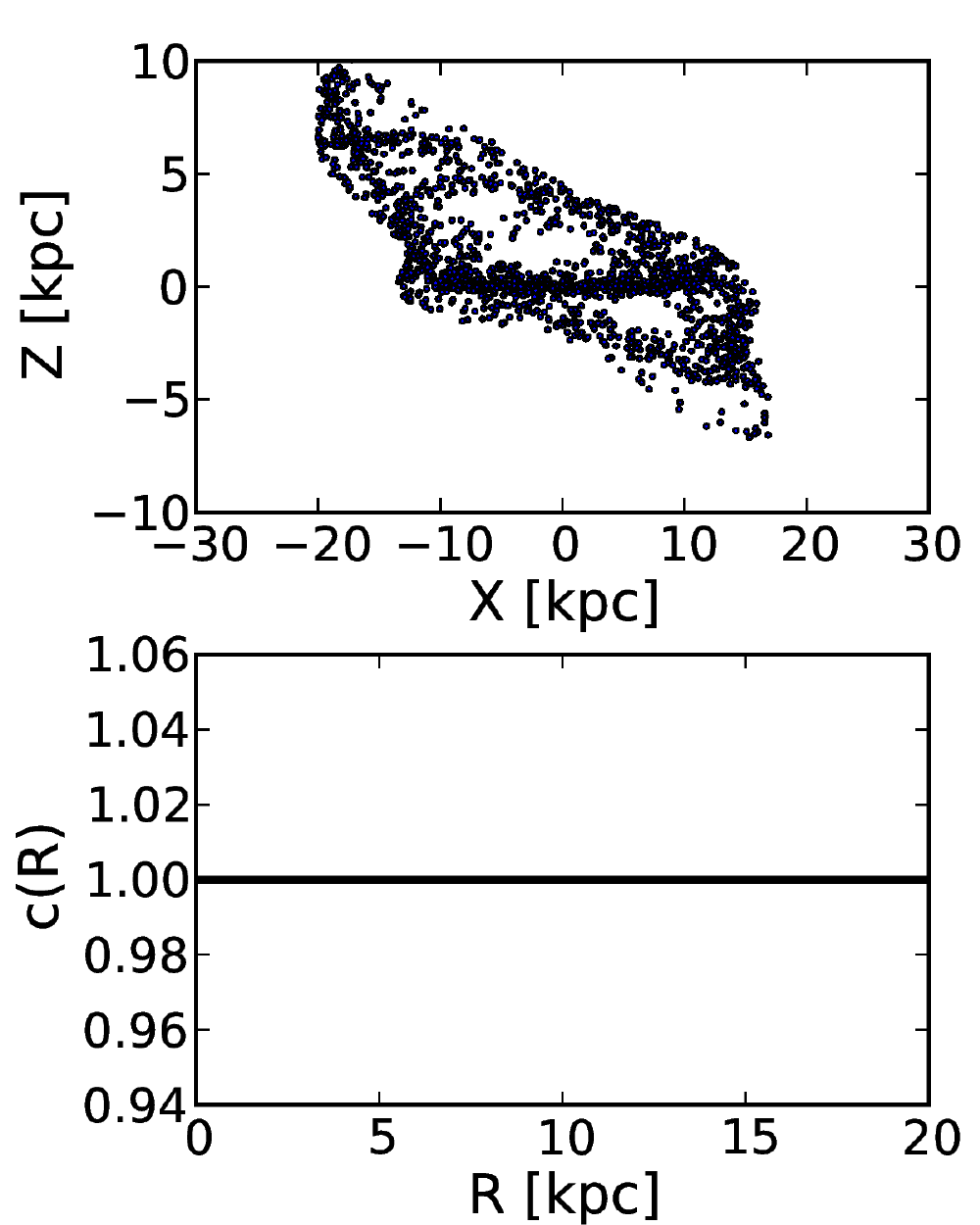}
\includegraphics[scale=0.54]{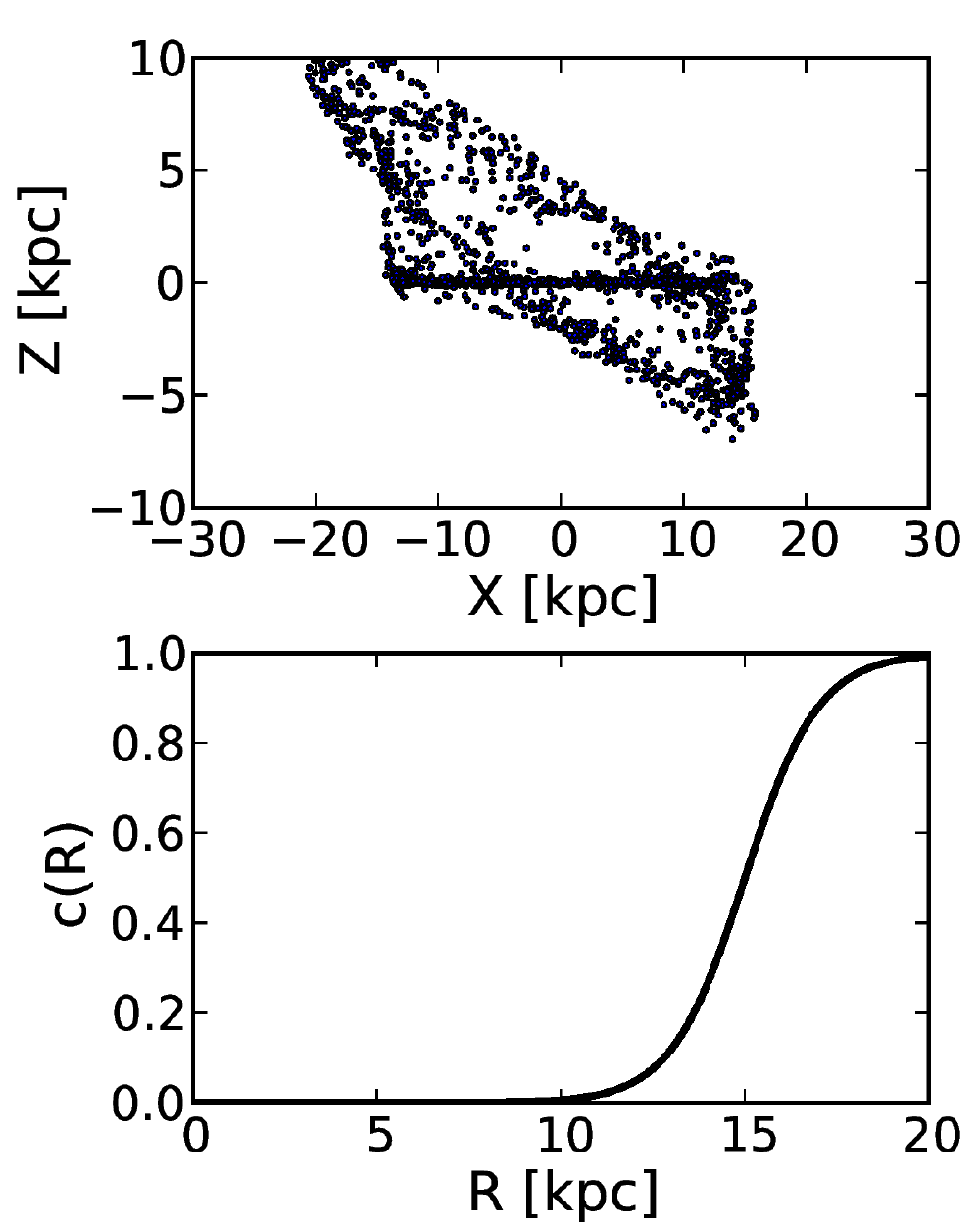}
\includegraphics[scale=0.54]{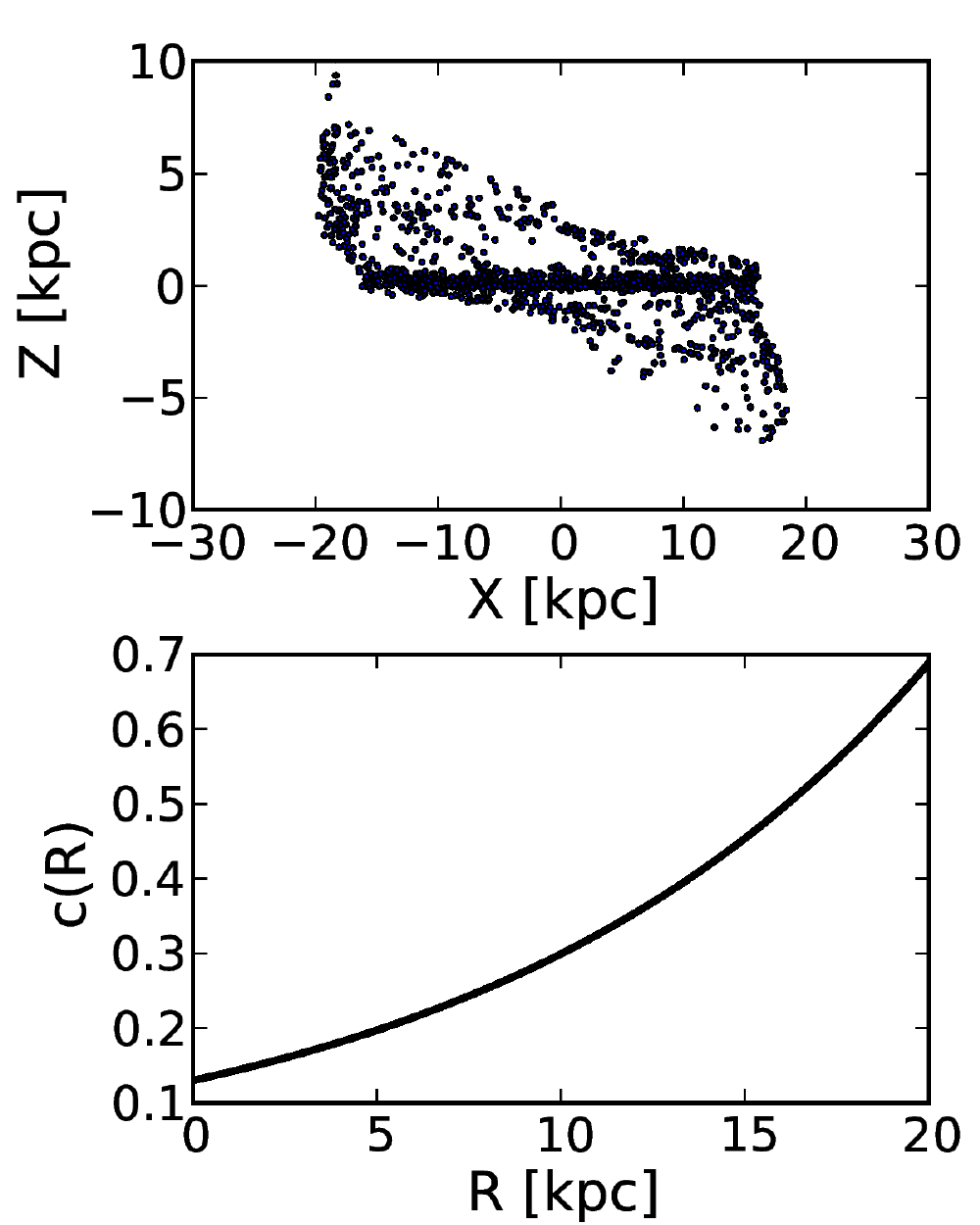}
\includegraphics[scale=0.32]{velwarp_90_uniform.pdf}
\includegraphics[scale=0.33]{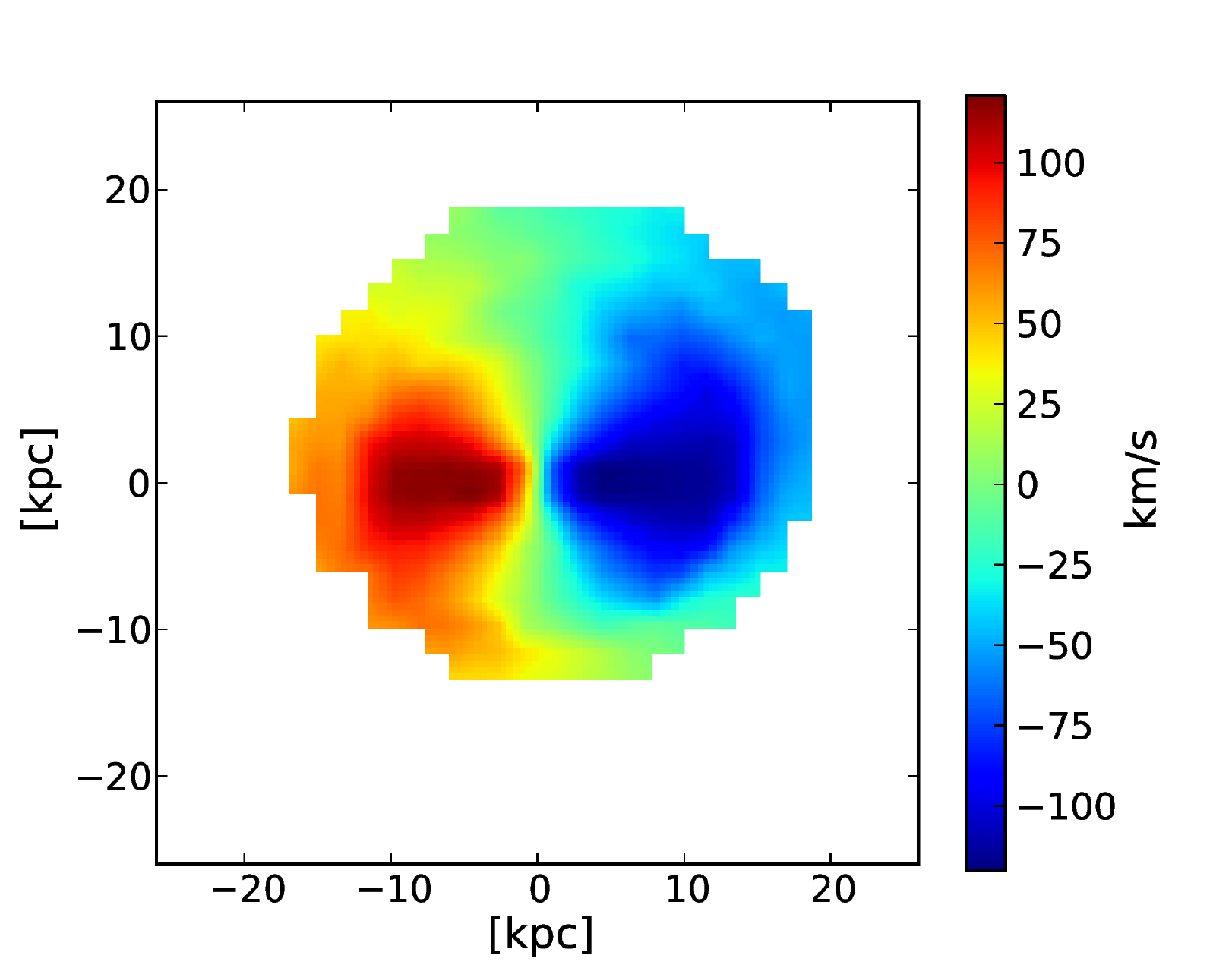}
\includegraphics[scale=0.33]{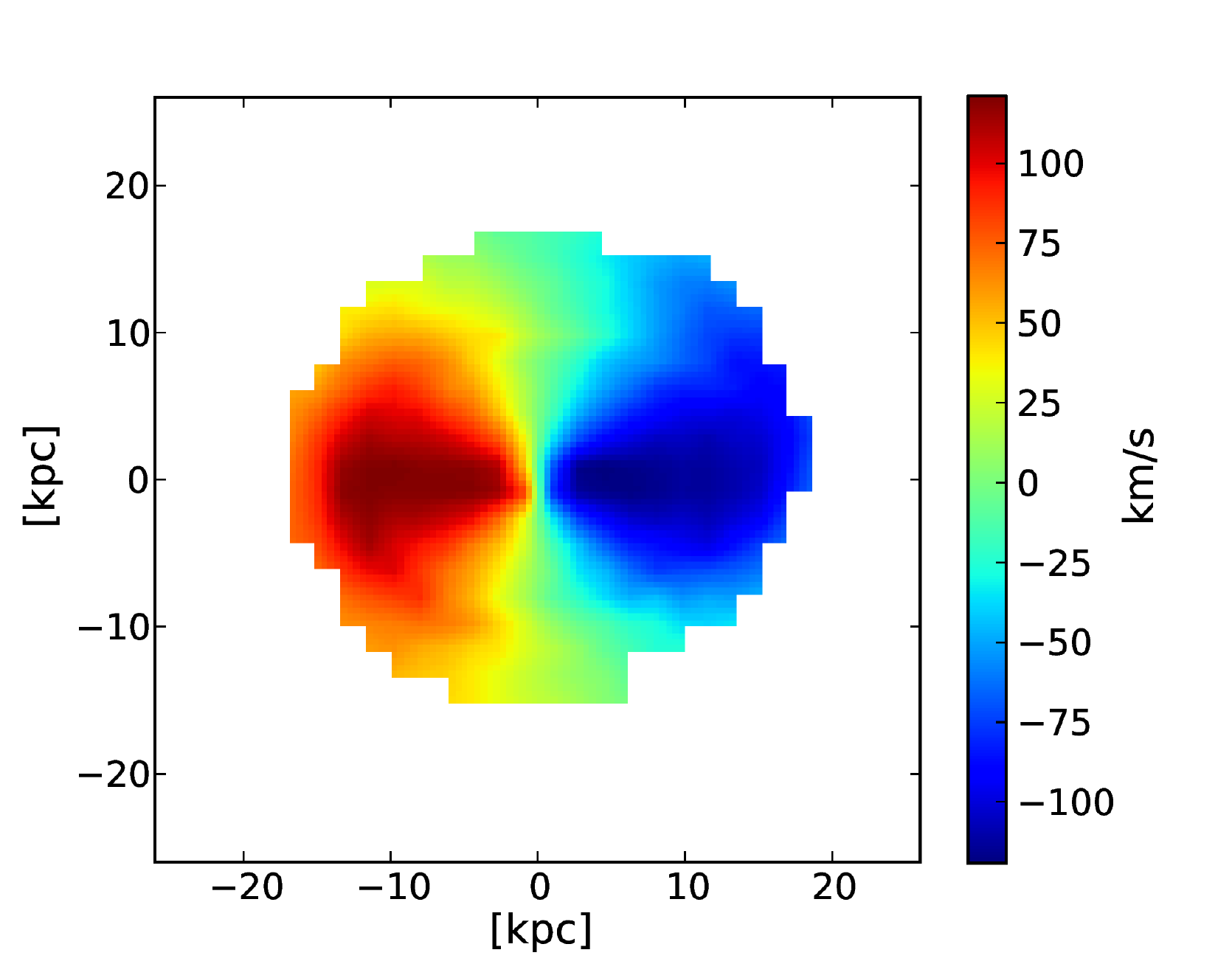}\\
\textbf{Figure~S3}: Comparison of the geometry of the final warped gas disk in the X-Z plane (top row) after 4~Gyrs under a moderate inclined ram wind for three different density profiles (see second row): uniform (left panel),  hyperbolic (middle panel), and exponential transition (right panel). The panels at the bottom display the corresponding velocity fields as observed on the sky with a disk inclination of 45~degree.
\end{center}
\end{figure*}

\clearpage


\begin{thebibliography}{}
\bibitem[Ade et al.(2013)]{ade13} Ade et al. 2013: 2013arXiv1303.5076P,\aap, submitted
\bibitem[Battaner \& Jimenez-Vicente(1998)]{Bat98} Battaner, E., \& Jimenez-Vicente, J.\ 1998, \aap, 332, 809 
\bibitem[Binney \& Tremaine(1987)]{Bin87} Binney, J., \& Tremaine, S.\ 1987, Princeton, NJ, Princeton University Press, 1987, 747 p., page 42--46
\bibitem[Briggs(1990)]{Bri90} Briggs, F.~H.\ 1990, \apj, 352, 15 
\bibitem[Dav{\'e} et al.(2010)]{dav10} Dav{\'e}, R., Oppenheimer, B.~D., Katz, N., Kollmeier, J.~A., 
\& Weinberg, D.~H.\ 2010, \mnras, 408, 2051 
\bibitem[Eke et al.(2005)]{eke05} Eke, V.~R., Baugh, C.~M., 
Cole, S., et al.\ 2005, \mnras, 362, 1233 
\bibitem[Gunn \& Gott(1972)]{Gun72} Gunn, J.~E., \& Gott, J.~R., III 1972, \apj, 176, 1 
\bibitem[Griv et al.(2002)]{Gri02} Griv, E., Gedalin, M., \& Yuan, C.\ 2002, \apjl, 580, L27
\bibitem[Haan \& Braun(2013)]{haa13} Haan, S., \& Braun, R., 2013, \mnras, submitted
\bibitem[Huang \& Carlberg(1997)]{Hua97} Huang, S., \& Carlberg, R.~G.\ 1997, \apj, 480, 503
\bibitem[Jiang \& Binney(1999)]{Jia99} Jiang, I.-G., \& Binney, J.\ 1999, \mnras, 303, L7 
\bibitem[J{\'o}zsa(2007)]{Joz07} J{\'o}zsa, G.~I.~G.\ 2007, \aap, 468, 903 
\bibitem[Kahn \& Woltjer(1959)]{Kah59} Kahn, F.~D., \& Woltjer, L.\ 1959, \apj, 130, 705 
\bibitem[Kamphuis \& Briggs(1992)]{Kam92} Kamphuis, J., \& Briggs, F.\ 1992, \aap, 253, 335 
\bibitem[Kamphuis et al.(2013)]{Kam13} Kamphuis, P., Rand, R.~J., J{\'o}zsa, G.~I.~G., et al.\ 2013, \mnras, 1790 
\bibitem[Kronberger et al.(2008)]{Kro08} Kronberger, T., Kapferer, W., Unterguggenberger, S., Schindler, S., \& Ziegler, B.~L.\ 2008, \aap, 483, 783 
\bibitem[L{\'o}pez-Corredoira et al.(2002)]{Lop02} L{\'o}pez-Corredoira, M., Betancort-Rijo, J., \& Beckman, J.~E.\ 2002, \aap, 386, 169 
\bibitem[L{\'o}pez-Corredoira et al.(2008)]{Lop08} L{\'o}pez-Corredoira, M., Florido, E., Betancort-Rijo, J., et al.\ 2008, \aap, 488, 511
\bibitem[L{\'o}pez-Corredoira \& Betancort-Rijo(2009)]{Lop09} L{\'o}pez-Corredoira, M., \& Betancort-Rijo, J.\ 2009, \aap, 493, L9 
\bibitem[Mayor  \& Vigroux(1981)]{May81} Mayor, M., \& Vigroux, L.\ 1981, \aap, 98, 1
\bibitem[Miyamoto \& Nagai(1975)]{Miy75} Miyamoto, M., \& Nagai, R.\ 1975, \pasj, 27, 533 
\bibitem[Navarro et al.(1996)]{nav96} Navarro, J.~F., Frenk, C.~S., \& White, S.~D.~M.\ 1996, \apj, 462, 563 
\bibitem[Nurmi et al.(2013)]{nur13} Nurmi, P., Hein{\"a}m{\"a}ki, P., Sepp, T., et al.\ 2013, \mnras, 2296 
\bibitem[Ostriker \& Binney(1989)]{Ost89} Ostriker, E.~C., \& Binney, J.~J.\ 1989, \mnras, 237, 785 
\bibitem[Revaz  \& Pfenniger(2001)]{Rev01} Revaz, Y., \& Pfenniger, D.\ 2001, Gas and Galaxy Evolution, 240, 278 
\bibitem[Revaz \& Pfenniger(2004)]{Rev04} Revaz, Y., \& Pfenniger, D.\ 2004, \aap, 425, 67 
\bibitem[S{\'a}nchez-Salcedo(2006)]{San06} S{\'a}nchez-Salcedo, F.~J.\ 2006, \mnras, 365, 555 
\bibitem[Tosa(1994)]{Tos94} Tosa, M.\ 1994, \apjl, 426, L81
\bibitem[van der Kruit(2007)]{Kru07} van der Kruit, P.~C.\ 2007, \aap, 466, 883
\bibitem[Weinberg(1998)]{Wei98} Weinberg, M.~D.\ 1998, 
\mnras, 299, 499 
\end{thebibliography}
\end{document}